\documentstyle[stwol,epsf]{article}
\epsfverbosetrue

\newcommand{\Dslash}{D\!\!\!\!\slash}

\def\Journal#1#2#3#4{{#1} {\bf #2}, #3 (#4)}

\def\NPB{{\em Nucl. Phys.} B}
\def\PLB{{\em Phys. Lett.}  B}
\def\PRL{\em Phys. Rev. Lett.}
\def\PRD{{\em Phys. Rev.} D}

\def\be{\begin{equation}}
\def\ee{\end{equation}}
\def\bea{\begin{eqnarray}}
\def\eea{\end{eqnarray}}

\begin{document}

\title{TOP QUARK THEORY}

\author{ S.~WILLENBROCK }

\address{Physics Department, U.~of Illinois, 1110 W.~Green St., 
Urbana, IL 61801, USA }

\twocolumn[\maketitle\abstracts{I review recent theoretical work on top-quark
physics within the standard model, beyond the standard model, and at 
the Planck scale.}]
 
In this talk I begin with a discussion of the top quark within the context
of the standard model, move on to consider top-quark physics beyond the 
standard model, and end with some speculative comments about the relation of 
the top quark to Planck-scale physics.

\section{Top within the Standard Model}

\indent There are only a few fundamental parameters associated with the 
top quark in
the standard model: the top-quark mass and the three CKM elements involving 
top. Below I discuss $m_t$ and $V_{tb}$; I will also discuss the top-quark 
cross section. 

\subsection{$m_t$}\label{subsec:mt}

\indent The world-average value of the top-quark mass from Run I at the 
Fermilab Tevatron
was reported at this conference to be 
\begin{equation}
m_t=175 \pm 6\;{\rm GeV} 
\end{equation}
based on roughly 100
$pb^{-1}$ of data.~\cite{Tipton}  It is appropriate to ask how precisely 
we desire to 
know $m_t$, and how precisely we can measure it.

This discussion is often 
carried out in the context of precision electroweak measurements.  The indirect
determination of the top-quark mass from these measurements is~\cite{Blondel}
\begin{equation}
m_t = 179 \pm 8 {+17 \atop -20} \;{\rm GeV} \;,
\label{eq:topmass}
\end{equation}
where the second uncertainty is obtained by varying the Higgs mass from 60 GeV
to 1 TeV.  The direct measurement of the top-quark mass is much more accurate 
than the indirect determination, so the measured mass can now be included in 
the precision studies to learn something about the Higgs mass (or, more 
generally, about electroweak symmetry breaking).  Rather than pursue this line
of thought further, I will leave it to another speaker,~\cite{Blondel} and 
take a different tack.

\begin{figure}[htb]
\begin{center}
\epsfxsize=0.49\textwidth
\leavevmode
\epsfbox[50 50 300 300]{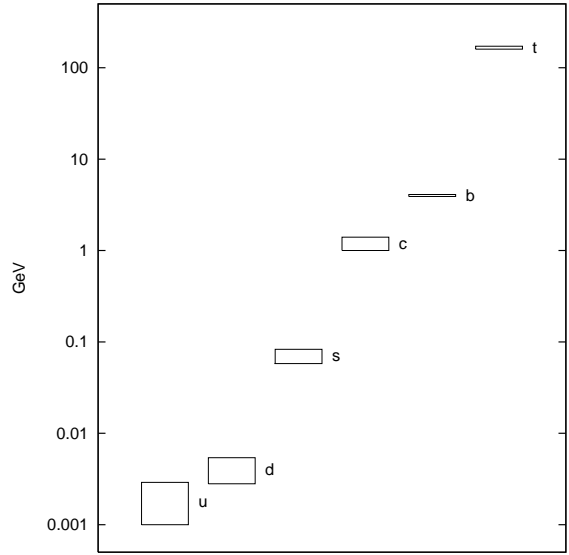} 
\end{center}
\caption{The quark mass spectrum.  The bands indicate the  
running $\overline{\rm MS}$ masses, evaluated at the quark mass 
(for $c,b,t$) or 
at 1 GeV (for $u,d,s$), and the associated uncertainty.}  
\label{fig:massspectrum}
\end{figure}

Figure 1 shows the quark mass spectrum on a logarithmic scale.  These are the 
running $\overline{\rm MS}$ masses, evaluated at the quark mass 
(for $c,b,t$) or 
at 1 GeV (for $u,d,s$).~\footnote{The light-quark masses are taken from a 
recent lattice calculation.~\cite{Gough,Bhattacharya} 
Note that they are significantly less than the Particle Data
Group values.~\cite{PDG}}  
The top-quark $\overline{\rm MS}$ mass is 
\begin{equation}
\overline{m}_t(\overline{m}_t)= 166 \pm 6\; {\rm GeV}\;.
\end{equation}
The percentage
uncertainty in the quark mass is proportional to the width of the band
in the figure.  It is 
evident that the larger the quark mass, the better known it is.  Since the 
top quark is the SU(2) partner of the bottom quark, one can imagine that we
will someday have a theory relating their masses, so it is desirable to know 
the top mass at least as well as we know the bottom mass.  

The most accurate determination
of the bottom mass comes from a comparison of the Upsilon mass with a 
lattice calculation,~\cite{Davies} yielding
\begin{equation}
\overline{m}_b(\overline{m}_b)=4.0 \pm 0.1 \; {\rm GeV} \;,
\end{equation}
an
uncertainty of $2.5\%$.  This corresponds to a measurement of the top mass to
4 GeV, which should be attainable in Run II at the Tevatron (2 $fb^{-1}$), 
based on roughly $10^3$ fully-reconstructed top events.  The uncertainty in 
the bottom mass is entirely theoretical, coming from lattice perturbation 
theory, and once this calculation is carried out to next order 
the uncertainty in 
the mass will be reduced to perhaps $1\%$.  This corresponds to a measurement
of the top mass to about 2 GeV, within reach of the Tevatron with 
10 $fb^{-1}$ of data (which requires additional running beyond 
Run II),~\cite{Tev2000}
and the CERN Large Hadron Collider (LHC).~\cite{ATLAS}  The LHC will 
be such a prolific source of top quarks, about a million fully-reconstructed 
events per year (at ``low'' luminosity, $10^{33}/cm^2/s$), that one can imagine
a measurement of the top mass to 1 GeV or even less.  The issue is entirely 
one of systematics, as the statistical uncertainty is negligible.

A more ambitious goal is to find a theory that relates one or more
quark masses to the 
gauge couplings.  I argue in the final section that the top quark is our
best candidate for such a relation to exist.  We know the electroweak gauge 
couplings with an 
accuracy of about $0.1\%$.  This corresponds to an uncertainty in the top 
mass of 200 MeV, which may be attainable with $e^+e^-$ and $\mu^+\mu^-$ 
colliders 
operating at the $t\bar t$ threshold.~\cite{NLC,muon}

Before leaving the topic of the top mass, I would like to mention an issue 
which must be resolved before Run II at the Tevatron.  It is important to 
understand hard gluon radiation in top events, and to correct for this 
effect when extracting the mass.  At present this effect introduces an 
uncertainty of 2.2 GeV in the mass.~\cite{Tipton}  The gluon radiation is
simulated with HERWIG.  However, it has been shown recently that HERWIG
produces too much gluon radiation in top events,~\cite{Orr1} 
despite the fact that it simulates gluon 
radiation from light quarks accurately.~\footnote{A previous version of 
HERWIG {\it underestimated} the amount of gluon radiation in 
top events;~\cite{Orr2} 
that turned out to be a bug, which has been fixed.}  This 
problem occurs in a region of phase space where perturbation theory is 
reliable, so it is not a consequence of multiple-gluon radiation.

\subsection{$V_{tb}$}\label{subsec:vtb}

\indent Of the three CKM matrix elements associated with the top quark, 
only $V_{tb}$
appears to have the potential to be measured directly.  Indirect measurements
of $V_{ts}$ and $V_{td}$ via $K$ and $B$ decays are discussed in another 
talk.~\cite{Gibbons}

If there are only three generations, $V_{tb}$ is already well known; in fact, 
it is the best-known CKM matrix element, although it has not been measured 
directly.  Three-generation unitarity, along with the known small 
values of $V_{cb}$ and $V_{ub}$, yields $V_{tb}= .9989-.9993$.~\cite{PDG}
Thus a direct measurement of $V_{tb}$ is only of real interest if we entertain
the possibility of more than three generations.  In this case, the value
of $V_{tb}$ is essentially unconstrained;
$V_{tb}= 0 - .9993$.~\cite{PDG}

\begin{figure}[htb]
\begin{center}
\epsfxsize=0.49\textwidth
\leavevmode
\epsfbox[48 530 410 660]{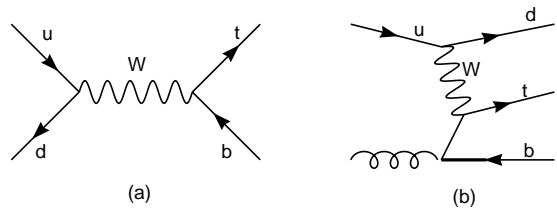} 
\end{center}
\caption{Single-top-quark production in hadron collisions: (a) quark-antiquark
annihilation, (b) $W$-gluon fusion.}  
\label{fig:singletop}
\end{figure}

The best way to measure $V_{tb}$ at a hadron collider is via single-top-quark
production, shown in Fig.~2.  There are two separate processes:
(a) quark-antiquark
annihilation,~\cite{Cortese} which is similar to the Drell-Yan process, and
(b) $W$-gluon fusion,~\cite{Dicus} which is similar to heavy-flavor production
via charged-current deep-inelastic scattering.
Both of these processes should be observed in Run II at the Tevatron.
They both involve the top-quark charged current, so their cross 
sections are proportional to $|V_{tb}|^2$.  The strategy is to measure the 
single-top cross section, which is given by 
\begin{equation}
\sigma=\sigma(t\bar b) BR(t\to bW)
\label{eq:tb}
\end{equation}
and to extract $BR(t\to bW)$ from $t\bar t$ production, given by 
\begin{equation}
\sigma=\sigma(t\bar t) [BR(t\to bW)]^2\;.
\label{eq:tt}
\end{equation}
One thus obtains $\sigma(t\bar b)$, which is proportional to 
$|V_{tb}|^2$.~\footnote{One cannot extract $V_{tb}$ from top decay alone.
For example, if $BR(t\to bW)$ is close to unity, one only learns that 
$V_{tb}>>V_{ts},V_{td}$.}  
This procedure requires a theoretical calculation of  
$\sigma(t\bar t)$, which is the topic of the next section.

Let's consider the strengths and weaknesses of the two separate single-top
processes.  The $q\bar q$ annihilation process has been calculated to 
next-to-leading order in QCD,~\cite{Smith} and much of the technology to 
extend this calculation to next-to-next-to-leading order 
exists.~\cite{Hamberg,Chetyrkin}  The $W$-gluon-fusion process has only been 
calculated to leading order, but the technology for the next-to-leading order 
calculation exists.~\cite{Laenen1,Giele}  The $q\bar q$ annihilation process
involves the quark distribution functions, which are better known than the 
gluon distribution function in the $W$-gluon-fusion process.  Furthermore,
the $q\bar q$ annihilation process benefits from its similarity to the 
Drell-Yan process, which can be used as a normalization. The 
$W$-gluon-fusion process has the advantage that it will be observable at the 
LHC, while $q\bar q$ annihilation will not, due to backgrounds.  The 
large rate of the $W$-gluon-fusion process at the LHC implies that 
the measurement of $V_{tb}$ will have negligible statistical uncertainty.  

$V_{tb}$ should be measured to about $10\%$ in Run II at the Tevatron,
and to about $5\%$ with additional data (10 $fb^{-1}$), via the $q\bar q$ 
annihilation process (assuming $V_{tb}$ is near unity), where the uncertainty
is statistical.  The measurement of 
$V_{tb}$ at the LHC via the $W$-gluon-fusion process will be limited mostly
by the uncertainty in the gluon distribution function: $\Delta V_{tb}\sim
\Delta g(x)/2$.  An uncertainty of $5\%$ requires knowledge of the gluon 
distribution function to $10\%$.  $V_{tb}$ can be extracted from the top width
measured from 
$e^+e^-$ and $\mu^+\mu^-$ colliders operating at the $t\bar t$ threshold:
$\Delta V_{tb}\sim \Delta\Gamma/2$.  An uncertainty in the width of less 
than $10\%$ may be possible,~\cite{NLC} yielding an uncertainty in $V_{tb}$ of
less than $5\%$.

The measurement of $V_{tb}$ at a hadron collider requires input from a 
variety of sources: deep-inelastic scattering (for the parton distribution 
functions), theory (for precise QCD calculations), and of course the 
actual experiment.  It's a good example of the coordinated effort that is 
often required to measure a fundamental parameter of the standard model.
 
\subsection{$\sigma(t\bar t)$}\label{subsec:sigma}

Top-quark production at the Tevatron is dominated
by quark-antiquark annihilation at moderate values of the parton momentum 
fraction, $x\sim 2m_t/\sqrt S \sim 0.2$, and thus should be calculable with 
high precision.
As we have just seen, an accurate calculation of the cross section for 
top-quark pair production is a necessary ingredient for the measurement of 
$V_{tb}$.  More importantly, this cross section is sensitive to new physics in
top-quark production and/or decay.  A new source of top quarks (such as 
gluino production, followed by the decay $\tilde g \to \tilde t t$) would 
appear as an enhancement of the cross section, and a new decay mode (such as 
$t \to \tilde t \tilde \chi^0$) would appear as a 
suppression.~\footnote{It has been
proposed that both of these processes are in fact present, and happen to 
compensate each other such that the observed cross section is close to the 
standard model.~\cite{Kane}}  Resonances in $t\bar t$ production would also
increase the top-quark cross section.~\cite{Hill1}

The top-quark cross section was calculated at next-to-leading order in QCD 
many years ago,~\cite{Nason} and was recently updated in Ref.~\cite{Catani}.
There are three sources of uncertainty in the calculation.  First, the value of
$\alpha_s(M_Z)$, when varied by $10\%$, results in a change in the cross
section of $6\%$.  Since the cross section is proportional to $\alpha_s^2$
(at tree level), it is perhaps surprising that the sensitivity of the cross 
section is not $20\%$.  The explanation is that there now exist parton 
distribution functions with a varying $\alpha_s(M_Z)$,~\cite{Martin,CTEQ} 
and it turns out that these largely compensate for the variation of the 
partonic cross section with $\alpha_s$.  Second, varying the parton 
distribution functions themselves (for fixed $\alpha_s(M_Z)$) leads to an
uncertainty of at least $3\%$.  This is judged by comparing the results using
the MRS~\cite{Martin} and CTEQ~\cite{CTEQ} parton distribution functions; 
however, these are best fits to some subset of the world's data, and are not
meant to represent the possible range of values for the parton distribution 
functions.  A set of parton distribution functions with built-in uncertainties
is badly needed.  Finally, the uncertainty due to the uncalculated higher-order
QCD correction is estimated by varying the factorization
and renormalization scales, and is about $7\%$.  Adding these three 
uncertainties in quadrature yields a total uncertainty of at least $10\%$.

A comparison of the theoretical cross section with the measured value, 
as a function of the top-quark mass, is given in Fig.~3.  The agreement is 
satisfactory at the one-sigma level, so there is no hint of new physics
in top-quark production at this time.

\begin{figure}[htb]
\begin{center}
\epsfxsize=0.48\textwidth
\leavevmode
\epsfbox[35 150 530 655]{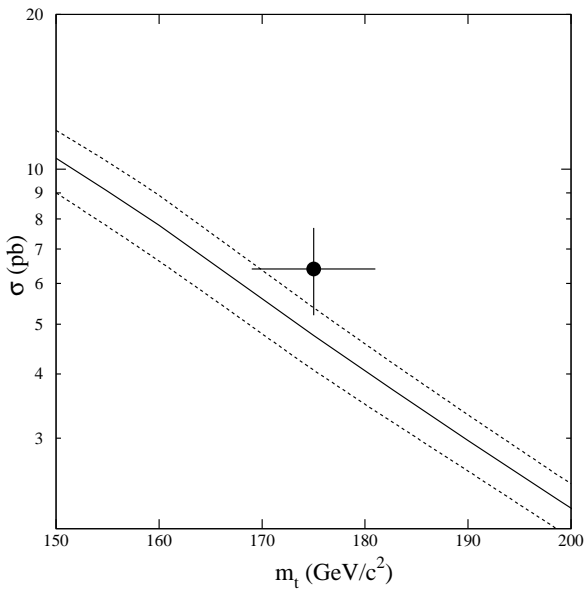} 
\end{center}
\caption[fake]{Cross section for $t\bar t$ production at the Tevatron 
{\it vs.} the
top-quark mass.  Theory curve, with error bands, is from the NLO QCD 
calculation of Ref.~\cite{Catani}.  The cross indicates the world-average 
mass and cross section.} 
\label{fig:topcross}
\end{figure}

Given that the uncertainties in $\alpha_s(M_Z)$ and the parton distribution 
functions will decrease with time (an uncertainty of $10\%$ in 
$\alpha_s(M_Z)$ is already conservative), the uncertainty due to the 
uncalculated higher-order QCD corrections will soon be dominant.  The 
next-to-next-to-leading-order QCD correction is needed to reduce the 
uncertainty 
further.  This is a technically challenging task, but the tremendous progress
in two-loop, and one-loop multi-leg, QCD calculations gives hope that this 
calculation will be possible.

There are attempts to sum, to all orders in $\alpha_s$, the contributions
from soft initial-state gluon emission.~\cite{Laenen2,Berger,Catani}  
This is relevant to all 
hadron-collider processes at large invariant mass, not just to top-quark 
production.  Single-gluon emission yields a correction proportional to 
$\alpha_s\ln^2 E_g$, where $E_g$ is the gluon energy.  This is a large 
correction for small $E_g$, but the integral over gluon energies is 
convergent, and yields a finite result.  Each additional gluon emission 
yields another such factor.  One finds that 
these terms approximately exponentiate, yielding a correction factor of
\begin{equation}
e^{\alpha_s \ln^2 E_g}\;.
\label{eq:sum}
\end{equation}
The integral over gluon energies is now divergent for small
$E_g$.~\footnote{A separate, unrelated, divergence occurs for small $E_g$
because the (running) $\alpha_s$
in the exponent is evaluated at the scale $E_g$.~\cite{Beneke}}  
This is not necessarily unphysical; it could be related to the fact that
QCD is nonperturbative for soft gluons.  Refs.~\cite{Laenen2,Berger}
take this point of view, and cut the integral off at the lower limit.
The result is that soft-gluon resummation increases the top-quark cross section
at the Tevatron by about $10\%$ above the next-to-leading-order cross section.
However, Ref.~\cite{Catani} argues that the divergence in the integral
is an artifact of the approximations made in deriving the correction 
factor of Eq.~\ref{eq:sum}.  An alternative summation procedure is 
proposed that does not diverge  
for small $E_g$.  This yields a correction due to 
soft-gluon emission of only about $1\%$ above the next-to-leading-order
cross section.  

\section{Top beyond the standard model}

Because top is so much heavier than the other fermions, it may be special in 
some sense.  There are many explicit models which implement this idea, too 
many to consider in this talk.  One such proposal, Topcolor,~\cite{Hill2} 
is discussed in another talk.~\cite{Lane}  Here I confine my remarks
to a model-independent observation which demonstrates the role that mass
might play in distinguishing the top quark from the other fermions.

Consider the gauge Lagrangian of a single quark doublet ($q_L=(u_L,d_L)$),
\begin{equation}
{\cal L} = i\bar q_L \Dslash q_L 
+ i\bar u_R \Dslash u_R + i\bar d_R \Dslash d_R
\label{eq:L}
\end{equation}
where the covariant derivatives contain the gauge fields.
The gauge interactions
possess an exact U(1)$\times$U(1)$\times$U(1) chiral symmetry, 
which independently
rotates the $q_L$, $u_R$, and $d_R$  fields by a phase. Quark mass
terms (generated by electroweak symmetry breaking) explicitly violate this 
symmetry, but because quark masses are small
compared to the weak scale (except for the top quark) we may regard the 
symmetry as being approximate.~\footnote{In the context of the standard Higgs 
model, the quark masses arise from (small) Yukawa couplings, which violate the 
chiral symmetry.}  
If a quark has non-standard interactions
with gauge bosons or with itself, these can be described by higher-dimension
operators in the Lagrangian.~\cite{Buchmuller} 
Some of these operators also explicitly violate
the chiral symmetry.  If the chiral symmetry is exact in the limit of 
vanishing quark mass, we expect such operators to be proportional to the quark
mass. For example, an anomalous magnetic-dipole interaction
term has the form
\begin{equation}
i\frac{m}{\Lambda^2} \bar q_L \sigma^{\mu\nu} q_R F_{\mu\nu} + H.~c.
\label{eq:dipole}
\end{equation}
where $\Lambda$ is the scale of the physics responsible for the 
non-standard interaction.  Since the top-quark mass is so much larger than
the other fermions, its anomalous magnetic-dipole interaction, if it exists,
could be much larger than that of the other quarks.

Because the top quark is so much heavier than the other fermions, it would
be a mistake to assume that its interactions are identical to those of the 
other fermions.  Among other things, top-quark interactions could involve
large CP violation and flavor-changing neutral currents.  
A large amount of phenomenological work has been done detailing the 
signatures of new physics in top-quark studies at present and future colliders,
and I cannot hope to review it all.  At best I can review the phenomenological
work which was presented at this conference.  

There were several contributions related to CP violation in top physics.
Ref.~\cite{Atwood} argues that the top quark is a sensitive probe of 
sources of CP violation beyond the standard model.  Ref.~\cite{Grzadkowski}
considers $e^+e^- \to t\bar t$, including CP violation in 
both production and decay.  Ref.~\cite{Rizzo} considers the same process with
the radiation of an additional gluon, as a probe of non-standard $gt\bar t$
(as well as $(\gamma,Z) t\bar t$) interactions.  Ref.~\cite{Gunion} considers
the process $e^+e^- \to t\bar t h$ to test the CP nature of the Higgs boson, 
$h$.  There was also a contribution on exploring the flavor-changing 
neutral current $Zt\bar c$ via $e^+e^- \to t\bar c$.~\cite{Han}  Finally,
an extended gauge model, based on the group 
SU(2)$_1\times$SU(2)$_2\times$U(1)$_Y$ was proposed in Ref.~\cite{Muller}.
The first SU(2) couples to the first two generations, and the second SU(2)
to the third generation.  The two SU(2) groups are spontaneously broken to 
the ordinary SU(2) of the standard model at a high scale.  The phenomenology
of the $W^{\prime},Z^{\prime}$ associated with the extra SU(2)
is considered.

To complement the phenomenology at future $e^+e^-$ colliders mentioned above,
I would like to spend some time on top phenomenology at hadron colliders,
a topic I feel has not received as much attention as it should.  It is well
known that top quarks can be polarized at an $e^+e^-$ collider by polarizing
the electron beam, and that this is a useful tool to study the weak decay
properties of the top quark, because the top quark decays before the strong
interaction has time to depolarize it.  There is an analogue of this tool at 
hadron 
colliders.  Although the top quarks are produced unpolarized in (unpolarized)
hadron collisions, the spins of the $t$ and $\bar t$ are 
correlated.~\cite{Kuhn,Mahlon,Stelzer,Brandenburg} 
(The spins are also correlated 
in unpolarized $e^+e^-$ collisions.~\cite{Parke}) 
This spin correlation can be used to study the weak decay properties of the top
quark by observing the angular correlations between the decay products of the 
$t$ and $\bar t$.  The spin correlation should be observed in Run II at the 
Tevatron.
 
Let's consider the origin of the spin correlation in the process $q\bar q\to 
t\bar t$, the dominant top-quark production process at the Tevatron.  
At energies large compared with the top mass, chirality conservation implies 
that the $t$ and $\bar t$ are produced with opposite helicities (``helicity
basis'').  At the 
other extreme, the $t$ and $\bar t$ are produced with zero orbital angular 
momentum at threshold, so spin is conserved.  Since the colliding quark
and antiquark have opposite spins (due to chirality conservation), the $t$ and 
$\bar t$ have opposite spins along the beam axis 
(``beamline basis'').~\cite{Mahlon}  Remarkably, there exists a basis which
interpolates at all energies between these two extremes (``diagonal basis''),
such that the $t$ and $\bar t$ spins are always opposite.~\cite{Parke}

The single top-quark processes discussed above are sources of polarized top
quarks at hadron colliders, since they involve the weak interaction.  Given
the large numbers of top-quark pairs and single top quarks that will be 
produced at the Tevatron and LHC, the spin correlation and the single-top
polarization should be powerful tools to analyze the properties of the
top quark.

\section{Top and the Planck scale}

There are many ideas relating the top quark to the Planck scale.  One of the 
prettiest is that the top quark is responsible for driving one of Higgs 
mass-squared 
parameters negative in supersymmetric grand-unified models, thus triggering 
electroweak symmetry breaking.  This is a feature of the 
minimal supersymmetric standard model, which is reviewed in another 
talk.~\cite{Ross}

Here I would like to consider attempts to understand the value of the top-quark
mass from first principles, something more ambitious than the minimal 
supersymmetric model, although compatible with it in some cases.  I first 
discuss two contributions to this conference, then end with my own personal 
favorite.

\subsection{Top and the Higgs potential}\label{subsec:Higgs}

The tree-level Higgs potential is given by the familiar expression
\begin{equation}
V(\phi) = -\mu^2\phi^2 + \lambda\phi^4
\end{equation}
where $\phi$ is the Higgs field.  The Higgs field acquires a 
vacuum-expectation value, $v\approx 250$ GeV, at the minimum of the potential.
The coupling $\lambda$ may be expressed
(at tree level) in terms of the Higgs mass and the vacuum-expectation
value by $\lambda =m_H^2/2v^2$.  At one
loop, the top quark renormalizes $\lambda$, making a negative contribution for
large Higgs-field values.
As a result, the Higgs potential can develop a second minimum, typically
at a value much greater than $v$, which has lower energy than the first
minimum.  To avoid this, the tree-level coupling $\lambda$ must be sufficiently
large that it is not driven negative by the one-loop top-quark contribution.
This leads to a lower bound on $\lambda$, and hence on $m_H$.  This is the 
well-known vacuum-stability bound on the Higgs mass, which was recently
updated in Ref.~\cite{Altarelli}.

The lower bound on the Higgs mass depends on the energy up to which 
the standard Higgs model is valid, because if the second minimum occurs at
an energy greater than this, it should not be taken seriously.  If one 
assumes that the standard Higgs model is valid up to the Planck scale, one 
obtains $m_H>135$ GeV.  However, if one trusts the model only up to 1 TeV,
one finds $m_H>70$ GeV, similar to the current experimental lower bound.

Froggatt and Nielsen~\cite{Froggatt} have argued that this second minimum 
should in fact be
degenerate with the first minimum, and that it should occur at the Planck 
scale.  These two conditions allow one to predict the Higgs mass and the 
top-quark mass.  They find $m_t=173\pm 4$ GeV, compatible with the 
experimental measurement, and $m_H=135\pm 9$ GeV.  They give a 
statistical-mechanics argument for the degeneracy of the vacuua based on the
coexistence of two phases at a first-order phase transition.

\subsection{Reduction of couplings}\label{subsec:reduction}

The couplings of a theory are in general unrelated to each other.  However, 
symmetries can relate couplings.  A well-known example is (supersymmetric) 
grand unification,
which relates the strong, weak, and hypercharge couplings by unifying them
into a single coupling, $g_U$, at the GUT scale.  This unification
is a consequence of the gauge symmetry of the grand-unified group.

The top-quark Yukawa coupling to the Higgs field is $y=\sqrt 2m_t/v\approx 1$,
a natural number, unlike the other fermions which have very small Yukawa 
couplings.  It is tempting to try to relate the top Yukawa coupling to the 
grand-unified gauge coupling, $g_U \approx 0.7$, since it also has a natural
value.  However, there is no way to relate a gauge coupling and
a Yukawa coupling via a symmetry in conventional field theory.~\footnote{An 
exception is $N\geq 2$ supersymmetric theories.}

If one attempts to simply relate the gauge and Yukawa couplings by imposing 
$y=\kappa g_U$, where $\kappa$ is a real number, one finds that in general
this relation is not preserved by renormalization; the (running) couplings 
depend on a scale, such that 
the relation can be true only at one particular scale.  
However, it sometimes occurs that for special values of $\kappa$, the relation
happens to be true at all scales, despite the fact that there is apparently
no symmetry
enforcing the relation.  Such a situation occurs in a (finite) SU(5) 
grand-unified model.~\cite{Kubo}  If one regards this special value of $\kappa$
as being favored by nature, one obtains a prediction for the top-quark mass of 
$m_t=183\pm 5$ GeV, an acceptable value.

\subsection{Top and string theory}\label{subsec:string}

I now come to my own personal favorite for Planck-scale physics, and the 
favorite of many others: string theory.  String theory is the leading
candidate for a quantum theory of gravity, and one can ask if it has anything
to say about the top-quark mass.  Since string theory is not a mature subject,
one cannot extract a definite prediction for the top-quark mass at this time,
but I will argue that such a prediction is foreseeable.

String theory has just one coupling, the so-called string coupling, and it
is naturally of order unity.
Thus the effective field theory which arises from string 
theory should have couplings which are all of order unity.  
This is encouraging because, as remarked in the previous section, the 
grand-unified coupling and the top-quark Yukawa coupling are both of order 
unity. 

Furthermore, string theory holds the promise of explaining why only the 
top-quark Yukawa coupling is of order unity, while all the other Yukawa 
couplings
are small.  The effective field theory which arises from string theory 
typically has a variety of discrete symmetries, which are the remnants of 
spontaneously-broken gauge symmetries (``discrete gauge symmetries'').
These discrete symmetries restrict the Yukawa couplings.  In models with 
three generations of quarks and leptons, one typically finds that there is at
most one nonvanishing Yukawa coupling, which we interpret as the top-quark
Yukawa coupling.  The other Yukawa couplings arise from higher-dimension 
operators in the effective field theory, and are suppressed by powers of the 
small ratio $<\phi>/M_{Planck}\sim 10^{-1}-10^{-2}$, where $<\phi>$ is the 
vacuum-expectation value of some scalar field.~\cite{Lopez,Faraggi,Chaudhuri}

\section{Conclusions}

The future holds increasingly-precise measurements of the fundamental 
parameters $m_t$ and $V_{tb}$.  However, I believe the study of top-quark 
physics promises much more.  I have argued that future study of the top quark
will yield either 
\begin{itemize}

\item New physics in the next generation of colliders

\item Information on Planck-scale physics

\end{itemize}
Either way, we learn something of great importance.

\section*{Acknowledgements}

I am grateful for conversations and correspondence with E.~Berger,
A.~Blondel, H.~Contopanagos, D.~R.~T.~Jones, A.~El-Khadra, E.~Laenen, 
P.~Langacker, R.~Leigh, T.~Liss, J.~Lykken, M.~Mangano, W.~Marciano,
S.~Parke, R.~Roser, Y.~Shadmi, T.~Stelzer, G.~Sterman, and P.~Tipton. 
This work was supported in part by Department of Energy grant 
DE-FG02-91ER40677.

\section*{Questions}
\noindent{\it J.~ Richman, University of California at Santa Barbara:}

You described various models that give predictions for 
the top-quark mass.  Do these models also provide information on the values of
CKM matrix elements?

\vskip 12pt
\noindent{\it S.~Willenbrock:}

The specific models I mentioned do not address the CKM matrix.  

\vskip 12pt
\noindent{\it H.~Nielsen, Niels Bohr Institute, Copenhagen:}

Let me answer the just put 
question for the case of our own work.  There is a very weak and indirect 
connection between the top-mass prediction and a fit we made for the other 
quark (and lepton) masses and the CKM matrix: the requirement of the degenerate
minima or degenerate vacuua we also used to predict the three fine structure 
constants using also a special gauge group hypothesis.  We used this gauge
group to a fermion mass-spectrum fit.

\vskip 12pt
\noindent{\it S.~Willenbrock:}

Thank you for reminding me of this work.~\cite{Froggatt2}

\vskip 12pt
\noindent{\it B.~Ward, University of Tennessee:}

In your discussion of the resummation 
results of Laenen {\it et al.} versus Catani {\it et al.}, did you mean to 
imply that the results of Catani {\it et al.} were the correct ones?

\vskip 12pt
\noindent{\it S.~Willenbrock:}

I certainly find the argument of Catani {\it et al.} compelling.
However, I believe this topic deserves further theoretical scrutiny.


\section*{References}
\begin{thebibliography}{99}
\bibitem{Tipton} P.~Tipton, these proceedings.

\bibitem{Blondel} A.~Blondel and S.~Pokorski, these proceedings.

\bibitem{Gough} B.~Gough {\it et al.}, hep-ph/9610223.

\bibitem{Bhattacharya} T.~Bhattacharya and R.~Gupta, hep-lat/9605039.

\bibitem{PDG} {\sl Review of Particle Physics}, Particle Data Group,
\Journal{\PRD}{54}{1}{1996}.

\bibitem{Davies} C.~Davies {\it et al.}, \Journal{\PRL}{73}{2654}{1994}.

\bibitem{Tev2000} {\sl Future Electroweak Physics at the Fermilab Tevatron:
Report of the tev\_2000 Study Group}, eds.~D.~Amidei and R.~Brock, 
FERMILAB-Pub-96/082 (1996).

\bibitem{ATLAS} ATLAS Technical Proposal, CERN/LHCC/94-43, LHCC/P2 (1994).

\bibitem{NLC} {\sl Physics and Technology of the Next Linear Collider},
NLC ZDR Design Group and the NLC Physics Working Group,
hep-ex/9605011.

\bibitem{muon} {\sl $\mu^+\mu^-$ Collider: A Feasibility Study},
$\mu^+\mu^-$ Collider Collaboration, BNL-52503 (1996).

\bibitem{Orr1} L.~Orr, T.~Stelzer, and W.~J.~Stirling, hep-ph/9609246.

\bibitem{Orr2} L.~Orr, T.~Stelzer, and W.~J.~Stirling, \Journal{\PLB}{354}{442}
{1995}.

\bibitem{Gibbons} L.~Gibbons, these proceedings.

\bibitem{Cortese} S.~Cortese and R.~Petronzio, \Journal{\PLB}{253}{494}{1991};
T.~Stelzer and S.~Willenbrock, \Journal{\PLB}{357}{125}{1996}.

\bibitem{Dicus} S.~Willenbrock and D.~Dicus, \Journal{\PRD}{34}{155}{1986};
C.-P.~Yuan, \Journal{\PRD}{41}{42}{1990}.

\bibitem{Smith} M.~Smith and S.~Willenbrock, hep-ph/9604223.

\bibitem{Hamberg} R.~Hamberg, W.~van Neerven, and T.~Matsuura, \Journal{\NPB}
{359}{343}{1991}.

\bibitem{Chetyrkin} K.~Chetyrkin, J.~K\"uhn, and M.~Steinhauser,
hep-ph/9606230.

\bibitem{Laenen1} E.~Laenen, S.~Riemersma, J.~Smith, and W.~van Neerven,
\Journal{\PRD}{49}{5753}{1994}.

\bibitem{Giele} W.~Giele, S.~Keller, and E.~Laenen, \Journal{\PLB}{372}{141}
{1996}.

\bibitem{Kane} G.~Kane and S.~Mrenna, hep-ph/9605351.

\bibitem{Hill1} C.~Hill and S.~Parke, \Journal{\PRD}{49}{4454}{1994};
E.~Eichten and K.~Lane, \Journal{\PLB}{327}{129}{1994}.

\bibitem{Nason} P.~Nason, S.~Dawson, and R.~K.~Ellis, \Journal{\NPB}{303}{607}
{1988}; W.~Beenakker, H.~Kuijf, W.~van Neerven, and J.~Smith, \Journal{\PRD}
{40}{54}{1989}.

\bibitem{Catani} S.~Catani, M.~Mangano, P.~Nason, and L.~Trentadue, 
\Journal{\PLB}{378}{329}{1996}; hep-ph/9604351.

\bibitem{Martin} A.~Martin, R.~Roberts, and W.~J.~Stirling,
\Journal{\PLB}{356}{89}{1995}.

\bibitem{CTEQ} CTEQ Collaboration, H.~Lai {\it et al.}, hep-ph/9606399.

\bibitem{Laenen2} E.~Laenen, J.~Smith, and W.~van Neerven, \Journal{\NPB}{369}
{543}{1992}.

\bibitem{Berger} E.~Berger and H.~Contopanagos, \Journal{\PLB}{361}{115}
{1995}; hep-ph/9603326.

\bibitem{Beneke} M.~Beneke and V.~Braun, \Journal{\NPB}{454}{253}{1995}.

\bibitem{Hill2} C.~Hill, \Journal{\PLB}{266}{419}{1991};
\Journal{\PLB}{345}{483}{1995}.

\bibitem{Lane} K.~Lane, these proceedings

\bibitem{Buchmuller} W.~Buchm\"uller and D.~Wyler, 
\Journal{\NPB}{268}{621}{1986};
G.~Gounaris, F.~Renard, and C.~Verzegnassi, \Journal{\PRD}{52}{451}{1995}.

\bibitem{Atwood} D.~Atwood, S.~Bar-Shalom, G.~Eilam, and A.~Soni,
Physics Reports (in progress).

\bibitem{Grzadkowski} B.~Grzadkowski and Z.~Hioki, hep-ph/9604301.

\bibitem{Rizzo} T.~Rizzo, in Ref.~\cite{NLC}.

\bibitem{Gunion} J.~Gunion, B.~Grzadkowski, and X.-G.~He, hep-ph/9605326.

\bibitem{Han} T.~Han and J.~Hewett, in progress.

\bibitem{Muller} D.~Muller and S.~Nandi, \Journal{\PLB}{383}{345}{1996}.

\bibitem{Kuhn} J.~K\"uhn, \Journal{\NPB}{237}{77}{1984}.

\bibitem{Mahlon} G.~Mahlon and S.~Parke, \Journal{\PRD}{53}{4886}{1996}.

\bibitem{Stelzer} T.~Stelzer and S.~Willenbrock, \Journal{\PLB}{374}{169}
{1996}.

\bibitem{Brandenburg} A.~Brandenburg, hep-ph/9603333. 

\bibitem{Parke} S.~Parke and Y.~Shadmi, hep-ph/9606419.

\bibitem{Ross} G.~Ross, these proceedings.

\bibitem{Altarelli} G.~Altarelli and G.~Isidori, \Journal{\PLB}{337}{141}
{1994}; J.~Casas, J.~Espinosa, and M.~Quiros, \Journal{\PLB}{342}{171}{1995}.

\bibitem{Froggatt} C.~Froggatt and H.~Nielsen, \Journal{\PLB}{368}{96}{1996}; 
hep-ph/9607302.

\bibitem{Kubo} J.~Kubo, M.~Mondrag\'on, M.~Olechowski, and G.~Zoupanos,
hep-ph/9606434; hep-ph/9512435.

\bibitem{Lopez} J.~Lopez and D.~Nanopoulos, \Journal{\NPB}{338}{73}{1990}.

\bibitem{Faraggi} A.~Faraggi, \Journal{\PLB}{274}{47}{1992};
B{\bf 278}, 131 (1992); \Journal{\NPB}{403}{101}{1993}.

\bibitem{Chaudhuri} S.~Chaudhuri, G.~Hockney, and J.~Lykken, 
\Journal{\NPB}{469}{357}{1996}.

\bibitem{Froggatt2} C.~Froggatt and H.~Nielsen, hep-ph/9607250.

\end{thebibliography}
\end{document}